# Prototype of a transient waveform recording ASIC

J. Qin,[a,b] L. Zhao,[a,b,1] B. Cheng,[a,b] H. Chen,[a,b] Y. Guo,[a,b] S. Liu,[a,b] and Q. An[a,b]

[a] *State Key Laboratory of Particle Detection and Electronics, University of Science and Technology of China, Hefei 230026, China*

[b] *Department of Modern Physics, University of Science and Technology of China, Hefei 230026, China*

  *E-mail*: zlei@ustc.edu.cn

ABSTRACT: The paper presents the design and measurement results of a transient waveform recording ASIC based on the Switched Capacitor Array (SCA) architecture. This 0.18 μm CMOS prototype device contains two channels and each channel employs a SCA of 128 samples deep, a 12-bit Wilkinson ADC and a serial data readout. A series of tests have been conducted and the results indicate that: a full 1 V signal voltage range is available, the input analog bandwidth is approximately 450 MHz and the sampling speed is adjustable from 0.076 to 3.2 Gsps (Gigabit Samples Per Second). For precision waveform timing extraction, careful calibration of timing intervals between samples is conducted to improve the timing resolution of such chips, and the timing precision of this ASIC is proved to be better than 15 ps RMS.

---

[1] Corresponding author.

# Contents

## 1. Introduction

Waveform digitization offers particle physics experiments the most direct and effective method for extracting event information. Traditionally, commercial flash Analogue-to-Digital conversions (ADCs) are employed for digitization [1-3]. However, the increase of channel number in large scale experiments precludes high speed ADCs, which suffer from high power dissipation, financial cost and system complexity. As an alternative approach, the SCA ASICs are becoming increasingly popular as a front-end readout solution for many triggered-event applications [4-7].

Over the last decade, several SCA ASICs have been reported in high energy physics experiments. For instance, the Swift Analogue Memory (SAM) with two channels and up to 2 Gsps sampling rate was designed to replace the Analog Ring Sampler (ARS) used for the ANTARES experiment [6]; the LABRADOR featuring nine channels, maximum sampling rate of 3.7 Gsps and on chip digitization was employed in the ANITA experiment [8-9]; the DRS4, the most wildly used SCA ASIC, was applied in the MEG [10], MAGIC [11] and other experiments because of its 6 Gsps sampling speed and low noise; and the PSEC4 used in large-area time-of-flight detector systems further pushes the sampling speed to over 15 Gsps and the bandwidth to 1.5 GHz [12].

Heavy ion research facility in Lanzhou (HIRFL) is the biggest heavy ion experimental facility in China, and the T0 detector, consisting of multiple multi-gap resistive plate chambers (MRPCs), is one of the key components in HIRFL CSR (Cooling Storage Ring) external target experiment. Precise time measurement is required for T0 detector with a time precision of 25 ps

RMS, and the current method uses NINO ASIC and FPGA TDC (Time to Digital converter) to get the time information. As an alternative solution, the waveform digitization method based on SCA architecture is proposed. In this paper, we describe a 128-cell two-channel sampling ASIC architecture and present the measurement performance.

## 2. Architecture

An overview of the major functional blocks of one channel in the ASIC is depicted in Figure 1. Each channel has a depth of 128 switched-capacitor sampling cells. The shared sampling clock is generated by an on-chip Delay-Locked Loop (DLL). During sampling, the write switches are controlled by sampling clocks form voltage controlled delay line (VCDL). Once a trigger signal is received by the ASIC, all switches turn off, and the captured waveform is stored in the capacitors. With the on-chip Wilkinson ADC, all sampled signals are digitized in parallel. Finally, the digitized data are serially read out using a shift register 'token' architecture. Further details of each block are described in the following sections.

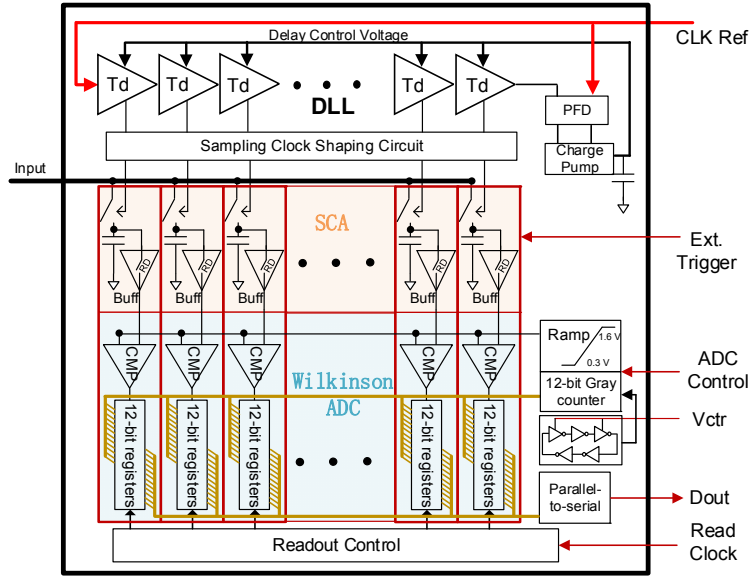

**Figure 1**. Block diagram of the SCA.

### 2.1 Timing generation

The sampling clock is generated by a 128-stage delay line, and each stage in the delay line is composed of two current-starved inverters and two normal inverters (Figure 2). Two serial current-starved inverters design ensures that the rising delay and falling delay are equal and less affected by the mismatch. The designed time delay per stage covers a range from 0.33 ns to 5 ns. An on-chip DLL circuit locks the sampling clock to an external reference clock at a dividing ratio of 128:1, and the sampling clock is almost independent of temperature, power supply and process because of the feedback operation mode. Considering the compromise between settling time and analog bandwidth, the pulse width of the sampling clock should be optimized carefully. Therefore, a pulse width reshape combinational logic circuit is designed to ensure only 4 sampling capacitors are connected to the input line at the same moment during sampling [13]. When an external clock with frequency $f_{in}$ (MHz) is fed into the chip, the sampling rate is boosted to 128· $f_{in}$ (MHz) automatically. Four cells connect to the input line at the same time, which means the settling time is $4 / (128 \cdot f_{in})$ (MHz$^{-1}$).

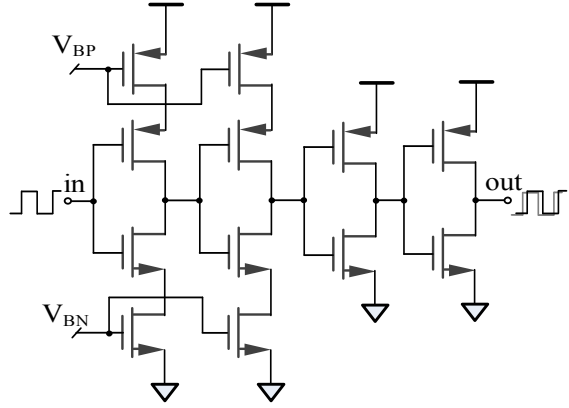

**Figure 2**. Schematic of the delay cell.

## 2.2 Sampling and Storage

Sampling circuit just contains a switch and a capacitor as shown in Figure 5, but the optimization of a fast sampling circuit is a multifaceted problem. The tradeoff among noise, bandwidth and distortion should be carefully made.

For high-precision time measurement, circuit noise will directly influence the final time meaurement resotluion.

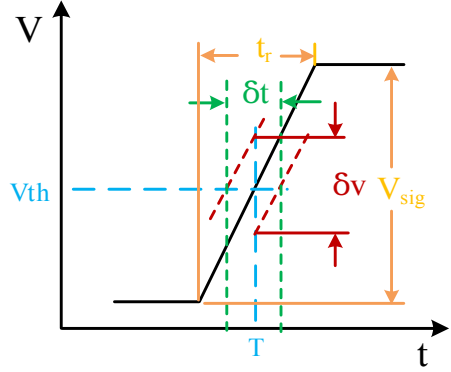

**Figure 3**. Time jitter caused by voltage noise.

The relationship between voltage noise (marked as $\delta v$) and time precision (marked as $\delta t$) is shown in Figure 3, where $V_{sig}$ is the signal voltage and $t_r$ is the rising time. We can obtain (detailed discussion can be found in [14]):

$$\frac{\delta v}{\delta t} = \frac{V_{sig}}{t_r} \tag{1}$$

$$\delta t = \frac{\delta v}{V_{sig}} \cdot t_r \cdot \frac{1}{\sqrt{n}} = \frac{\delta v}{V_{sig}} \cdot \frac{t_r}{\sqrt{t_r \cdot f_s}} = \frac{\delta v}{V_{sig}} \cdot \frac{1}{\sqrt{3 f_s \cdot f_{3dB}}} \tag{2}$$

where $n$ is the number of sample points lying on the leading edge. When a polynomial fitting with these points is applied to the leading edge, the time uncertainty is reduced by $\sqrt{n}$. The value of $n$ also can be expressed as $t_r \cdot f_s$ and $f_s$ is the sampling speed while the rise time $t_r$ mainly is determined by the analogue bandwidth of the SCA as $t_r \approx 1/(3 f_{3dB})$.

According to (2), we can estimate the time uncertainty caused by voltage noise. Besides, the effect of no-uniform sampling intervals and power ripple also contribute to the final time uncertainty. To satisfy the 25 ps requirement, the time uncertainty caused by voltage noise should be controlled under 10 ps. In our application, the minimum signal voltage $V_{sig}$ the SCA received is about 20 mV, $f_s$ is 2 Gsps, $f_{3dB}$ is about 450 MHz, so the voltage noise should be no more than 0.32 mV.

For an S/H circuit, the sampling resolution is limited by the integrated noise expressed as

$$v_{rms} = \sqrt{\frac{kT}{C}} \tag{3}$$

where k is Boltzmann's constant and T is temperature in Kelvin. To match the 12-bit voltage resolution of 1 V dynamic range, the noise should be controlled under 0.25 mV, and a reference capacitor value of 78 $fF$ is needed. Actually, a 110 $fF$ PMOS transistor gate capacitor is implemented, because increasing the capacitor properly makes switch charge injection and leakage current effects less prominent.

All sampling switches connect to the input signal line, and the parasitic capacitors contributed by sampling switches dominate the load capacitance. So reducing the size of the sampling switches increases the analogue bandwidth. However, a small switch exhibits high on-resistance and bad on-resistance flatness versus input voltage, which deteriorates the performance. The reason is that, as we know, RC circuit, consisted by a switch and a sampling capacitor, causes time delay to the input signal, if the on-resistance varies with the input voltage, then the signal delay varies with the voltage level, which produces signal distortion. Finally, an optimized complementary switch is adopted. The simulated on-resistance value of the complementary switch is depicted in Figure 4.

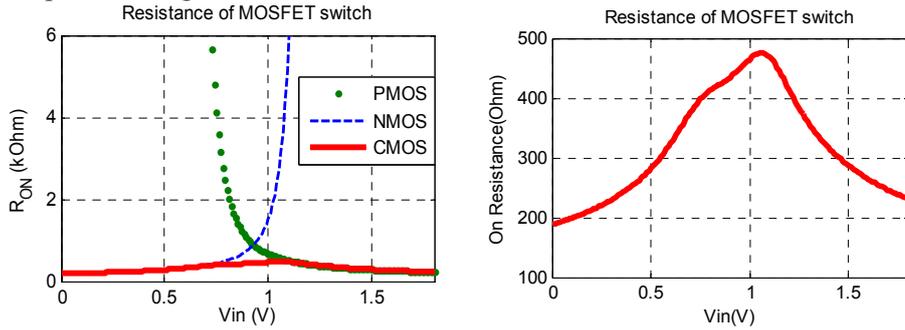

**Figure 4**. On-resistance of the complementary switch.

## 2.3 Digitization and Readout

Digital quantization of sampled values are done by Wilkinson ADC because of its very good parallelism integration [12]. And each sample cell has a dedicated comparator and a 12-bit latch as shown in Figure 5. When the digitization process is started, a voltage ramp and the stored voltage are put at the inputs of a comparator. In the meanwhile, the 12-bit gray counter begin counting. Once the ramp value excesses the stored voltage, the comparator fires and latches the counter value for readout.

The shared ramp signal is generated by using a current source and an internal capacitor, the 12-bit gray counter should run at a specific frequency to match ramping rise time, and a 5-stage voltage controlled ring oscillator provides the fast counter clock, which is adjustable between 400 MHz and 1.7 GHz.

The latched counter bits of each cell are driven to the bus under the control of a shift register 'token' signal, generated by a read control model (Figure 1). The 12-bit parallel data are fed into a serializer, and the serial data is read out to the FPGA at a rate of 200 MHz.

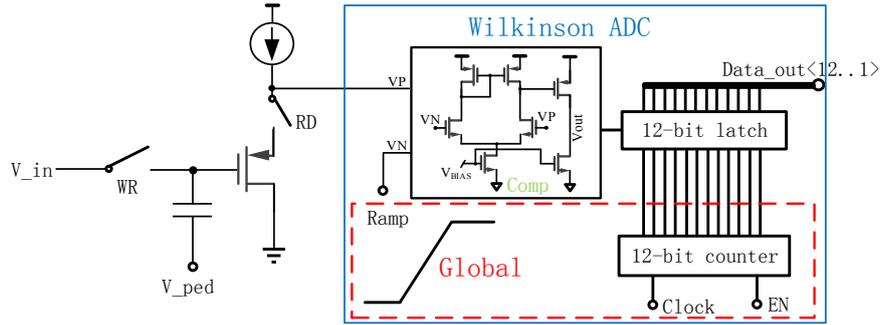

**Figure 5**. Simplified schematic of a single cell. The Wilkinson ADC is outlined by the rectangle.

## 3. Performance

Tests of the ASIC performance have been carried out using a custom evaluation board, and the test system is shown in Figure 6. We used a signal source (ROHDE&SCHWARZ SMA 100A) to provide high quality sinusoidal signals. A Xilinx Spatan-6 FPGA based evaluation board was used for chip configuration and data processing.

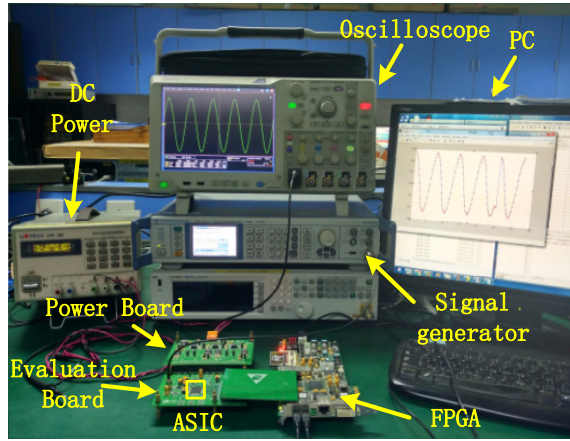

**Figure 6**. ASIC test system.

In this section, we present measurements of the sampling speed, DC transfer function, bandwidth, AC response and fast pulse timing, as well as calibration of sampling intervals.

### 3.1 Sampling speed

The sampling delay is controlled by an analogue voltage, which is set by the on-chip DLL automatically. In order to verify the performance of the sampling circuit, the analogue control voltage is drawn out of the chip. We change the reference input clock frequency and observe the variation of the voltage. The sampling rate as a function of the control voltage is depicted in Figure 7. A stable sampling rate ranging from 76 Msps to 3.2 Gsps is available and an agreement is shown between test values and SPICE simulation results. We set the sampling rate at a constant value of 2 Gsps in the following tests.

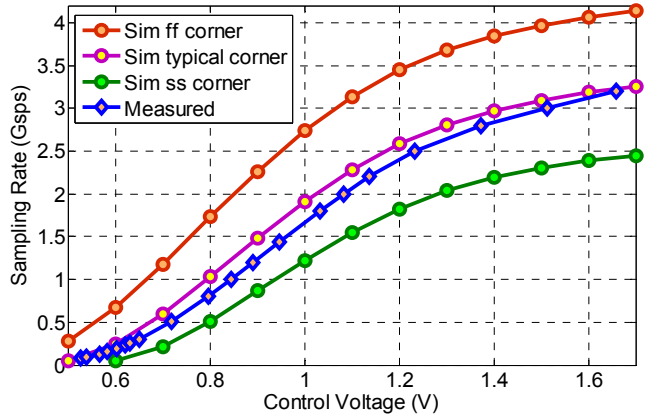

**Figure 7**. Sampling rate as a function of control voltage.

### 3.2 DC transfer function

The DC transfer function (measured ADC counts as a function of input signal amplitude) depends on configuration of the Wilkinson ADCs. We set the 12-bit gray counter clock at 1 GHz to reduce the conversion time as much as possible, and adjust the ramp charging current to maximize the use of ADC range and get a better resolution. Figure 8 shows the DC transfer function for a single channel. The 1 V input range spans 300-4000 ADC counts, corresponding to 3700 counts (11.85 bits) of resolution. After voltage calibration, the DC noise is measured to be ~ 1.4 mV, and the effective dynamic range is 9.5 bits. An integral nonlinearity (INL) of better than 2% is shown for the input range.

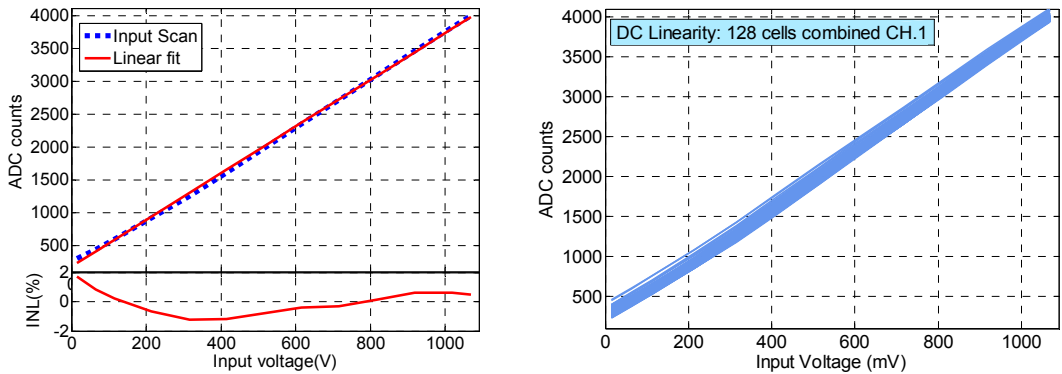

**Figure 8**. The DC transfer function for a single channel. (L) The data are the mean of all 128 cells. The linear fitting curve presents a lope of 3.55 counts/mV. (R) Raw DC scans for all 128 cells.

Based on the DC transfer function, the voltage calibration can be done to eliminate the difference of offsets and gains spread in cells caused by the process. And the linear interpolation method is applied to all the storage cells.

### 3.3 Response to sinusoids

The AC response was evaluated using sinusoidal signals with the voltage calibration mentioned in the section 3.2. Figure 9 shows a 51 MHz sine signal recorded by the ASIC after voltage calibration versus an oscilloscope (Tektronix DPO 5104) with a sampling rate of 2.5 Gsps, and the waveform recorded by the ASIC is much smoother than that of the oscilloscope.

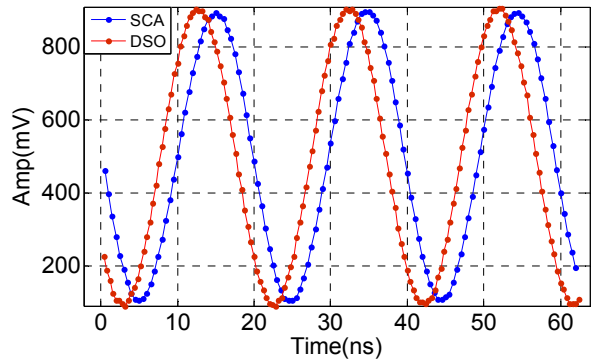

**Figure 9**. A 51 MHz sinusoidal signal recorded by ASIC after voltage calibration vs an oscilloscope with a sampling rate of 2.5 Gsps.

For a given frequency sinusoidal signal, by varying the input amplitude and comparing to the measured amplitude, the "AC transfer function" is determined as shown in Figure 10. Saturation is observed for 200 MHz signals at large amplitude obviously. This is not only an effect due to finite bandwidth, but also because of the insufficient slew rate. And it shows that a linear input range of 800 mV is available within 200 MHz.

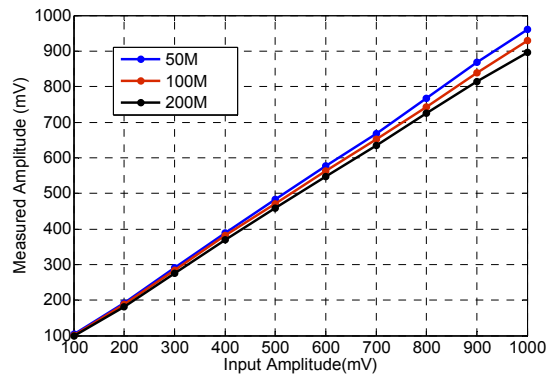

**Figure 10**. Measured sinusoid amplitude, as a function of input amplitude.

**3.4 Bandwidth**

Many factors influence the analog bandwidth, which include the off-chip circuitry such as an analog buffer amplifier, board and packaging parasitics, etc. and chip-level parasitic resistances and capacitances. Here, we only pay attention to the latter. Each cell has a parasitic resistance of the input line and a parasitic capacitance contributed by input line and sampling switch, and they cause a amplitude attenuation to the input signal.

As same as the "AC transfer function" measurement, the bandwidth was also conducted by comparing the amplitude of input sinusoids to the amplitude of waveforms recorded by the ASIC. For the SCA architecture, there will be an amplitude attenuation along the chip input line because of the series parasitic resistances and capacitances, and the test result in Figure 11(L) verifies the analysis. So we selected the amplitude of last cell to calculate the bandwidth. A range of sinusoidal input from 50 MHz to 800 MHz with the unified amplitude were generated by a signal source (ROHDE&SCHWARZ SMA 100A). Figure 11(R) shows the measured attenuation of the ASIC as a function of input frequency. And the analog bandwidth achieved is about 450 MHz.

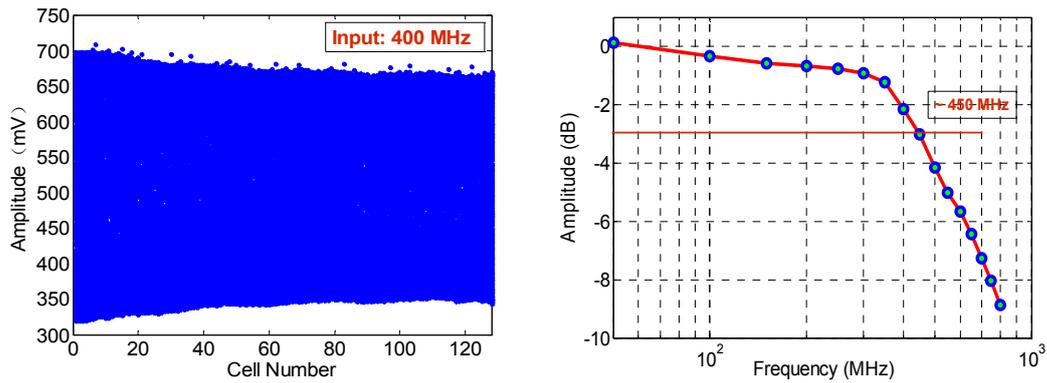

**Figure 11**. (L) Overlaying sine waves on 128 cells. (R) Frequency response of the ASIC.

### 3.5 Sampling intervals

SCA ASICs inevitably exhibit the non-uniform sampling intervals issue because of the process variations, so careful calibration is required to improve the timing resolution of such systems. The 'zero-crossing' time-interval calibration method is employed to find the intervals variations and compensate the waveform samples [14-15].

In the test, a 50.3 MHz sine wave is fed into the ASIC. Figure 12(L) depicts the distribution of the sampling intervals after 'local' time calibration and 'global' time calibration, and Figure 12(R) illustrates the nonlinearities (INL and DNL). The small time interval at the last sample bin is caused by a fixed DLL lock-error when wrapping the sampling from the last to the first.

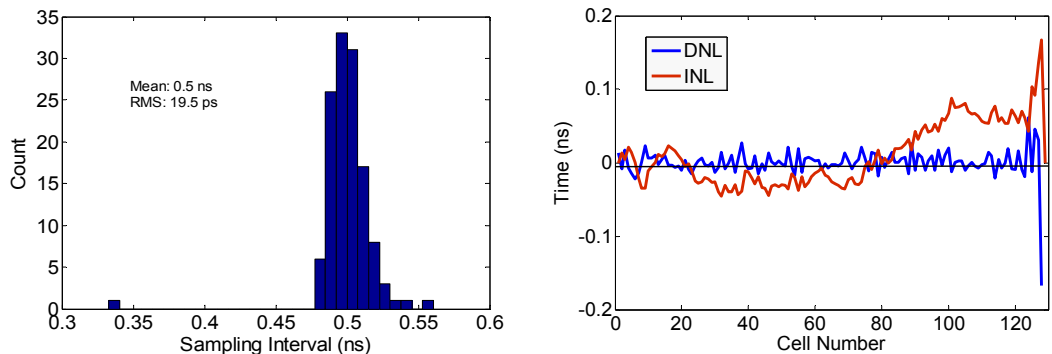

**Figure 12**. Sampling intervals of 128 cells. (L) A histogram of the intervals. (R) The differential (DNL) and integral non-linearity (INL) of intervals.

### 3.6 Waveform timing

We evaluated the waveform timing precision by measuring the time difference between two signals. The detailed procedure is as follows, a fast pulse generated by a function generator (AFG3252) is split into two branches, which are connect to two channels via two different length cables, and the time difference between the arrivals is measured.

Two channels captured the waveforms at the same time, and typical waveforms after voltage calibration are shown in Figure 13(a). A 6-order polynomial functional fitting is performed to the leading edge of each pulse. A global threshold of 400 mV was used to get the arriving times, and the difference between the arriving times was calculated. Since two channels are used to determine this difference, and the two channels are identical and independent, so the timing

– 8 –

precision per channel is $1/\sqrt{2}$ of the standard deviation of the time differences. Figure 13(b) shows the histogram of raw time difference distribution of nearly 2000 times of measurements and Figure 13(c) shows the result with time intervals calibration. The cable delay is found to be 14.33 ns, and the timing precision per channel is improved from 48.09 ps to 14.50 ps RMS after time intervals calibration.

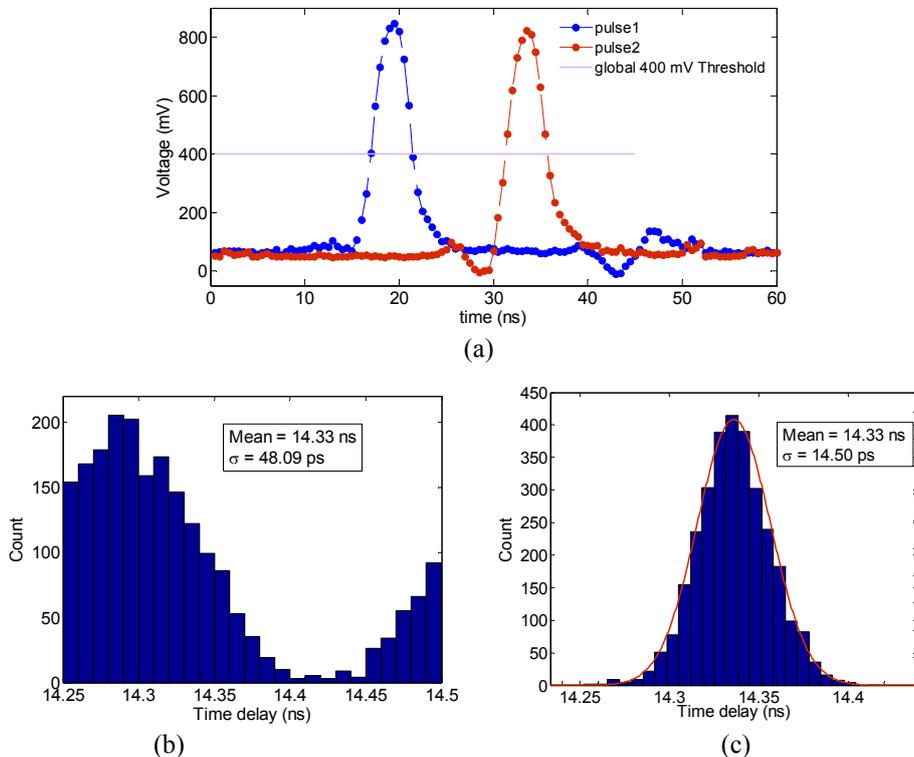

**Figure 13**. (a) Split pulses recorded by two channels. (b) and (c) are with and without calibration of non-uniform sampling intervals.

## 4. Conclusion

A SCA ASIC device has been designed and fabricated in a 0.18 μm CMOS process with an analog bandwidth of about 450 MHz, capable of being sampled stably at 2 Gsps. With sampling intervals calibration and rising-edge fitting, it is possible to extract precision timing information with a timing resolution about 15 ps RMS.

**Acknowledgments**

The authors would like to thank Wei Wei in Institute of High Energy, CAS for his constant help in our ASIC design. We also appreciate the discussion during the previous SCA design work with Wei Wei, Zhi Deng of Department of Engineering Physics in Tsinghua University and others, which enhanced our understanding about SCA. We thank Fukun Tang of Enrico Fermi Institute in University of Chicago for his invaluable guidance in the previous SCA design work.

This work was supported in part by Key Research Program of Frontier Sciences, CAS under Grant QYZDB-SSW-SLH002, in part by Science and Technological Fund of Anhui Province for Outstanding Youth under Grant 1708085J07, in party by the Knowledge Innovation Program of the Chinese Academy of Sciences under Grant KJCX2-YW-N27, in part by the Fundamental

Research Funds for the Central Universities under Grant WK2360000001, and in part by the CAS Center for Excellence in Particle Physics (CCEPP).